\documentclass[12pt,a4paper]{article}
\usepackage{amsmath,amssymb, amsfonts, amsthm, graphics, graphicx}

\title{Quantum Non-Objectivity from Performativity of Quantum Phenomena}

\author{Andrei Khrennikov\\ International Center for Mathematical Modeling\\
 in Physics, Engineering, Economics, and Cognitive Science\\
Linnaeus University, S-35195, V\"axj\"o-Kalmar, Sweden\\
Andrew Schumann\\ 
Department of Cognitivistics \\University of Information Technology and Management\\
 Sucharskiego 2, 35-225, Rzeszow, Poland}

\begin{document}
\maketitle

\begin{abstract}
We analyze the logical foundations of quantum mechanics (QM) by
stressing non-objectivity of quantum observables which is a
consequence of  the absence of logical atoms in QM.  We argue that
the matter of quantum non-objectivity is that, on the one hand,
the formalism of QM constructed as a mathematical theory is
self-consistent, but, on the other hand, quantum phenomena as
results of experimenter's performances are not self-consistent.
This self-inconsistency is an effect of that the language of QM
differs much from the language of human performances. The first is
the language of a mathematical theory which uses some Aristotelian
and Russellian assumptions (e.g., the assumption that there are
logical atoms). The second language consists of performative
propositions which are self-inconsistent only from the viewpoint
of conventional mathematical theory, but they satisfy another
logic which is non-Aristotelian. Hence, the representation of
quantum reality in linguistic terms may be different: from a
mathematical theory to a logic of performative propositions. To
solve quantum self-inconsistency, we apply the formalism of
non-classical self-referent logics.
\end{abstract}

keywords: non-objectivity of quantum observables, logical
structure of quantum description, self-inconsistency,
self-referent logic, photon existence, Grangier-type experiments,
coefficient of second order coherence, prequantum classical
statistical field theory, \emph{Physarum polycephalum}

\section{Introduction}

At many occasions, Niels Bohr repeated that quantum mechanics (QM)
does not yield a description of objective reality; in particular,
the values of quantum observables cannot be assigned before
measurement (they are not properties of objects) \cite{BR}:
{\small ``There is no quantum world. There is only an abstract
quantum physical description. It is wrong to think that the task
of physics is to find out how Nature is. Physics concerns what we
can say about Nature.''}\footnote{As is typical for the Bohr's writings, the
meaning of this statement is not clear. Did he deny the reality of
quantum systems -- atoms, electrons, photons?  (The discussion in
the present paper is essentially about the nature of photon. Here
we remark that initially Bohr was critical to Einstein's idea
about quanta of the electromagnetic field. However, after the
1920s he, in fact, accepted Einstein's idea.)}

Non-objectivity of quantum observables\footnote{Here we have to be
very careful with the terminological issue. Bohr definitely did
not consider quantum observables as properties of objects, quantum
systems. Thus, from this point of view they are non-objective. At
the same time, since for him measurement is performed by classical
macroscopic devices, the result of measurement is objective as the
output of a classical device. This problem, whether Bohr is
for/contra realism, was analyzed in very detail by A. Plotnitsky,
see \cite{P3}, \cite{P4}.} has tremendous consequences for the
physical picture of micro-phenomena. It is not easy (if possible
at all) to imagine lawful nature without objective properties of
physical systems. Therefore, the idea that non-objectivity implies
that at the microlevel the universe is totally lawless is a very
common consequence of non-objectivity. Although the interpretation
of the quantum universe as the totally lawless universe is very
popular \footnote{Among the most active advertisers of this
picture, we can mention, e.g., Anton Zeilinger \cite{Z1},
\cite{Z2}, whose theoretical considerations are supported by the
incredible experimental research in quantum foundations. He and
Caslav Brukner wrote a series of papers \cite{B1}--\cite{BX} on
{\it irreducible quantum randomness}. (We remark that the idea
that quantum randomness differs crucially from classical
randomness was discussed already by von Neumann\cite{VN}.)}, many
experts in quantum foundations including even ``quantum
orthodoxes'', i.e., those who can not even imagine to go beyond
quantum theory, feel unsatisfactoriness by appealing to the
picture of the lawless universe for so lawful formalism as QM.
Unfortunately, the only way to escape lawlessness is to appeal to
quantum nonlocality: to claim that quantum observables are
objective, but there is action at a distance. At the same time by
the aforementioned reason, i.e., unwilling to go beyond QM,
``quantum orthodoxes'' do not like Bohmian mechanics. This
situation is definitely self-contradictory.

First of all, we repeat our arguments from \cite{IJTP_KHR}
supporting non-objectivity of quantum observables. Theoretically,
the origin of non-objectivity was well explained by Bohr who
pointed that the contribution of a measurement device into the
result of measurement is {\it irreducible}.\footnote{It is very
common to speak about {\it irreducible quantum randomness}
\cite{VN}, \cite{Z1}--\cite{BX}. However, it is very difficult, if
possible at all, to define irreducible randomness in mathematical
terms.} Moreover, recent experiments on {\it quantum
contextuality}, see \cite{Hasegawa}, can be definitely interpreted
as supporting non-objectivity of quantum observables; in any event
there is no even a trace of nonlocality.  Thus, if one is not
really addicted on nonlocality, one cannot ignore that
non-objectivity is the most fundamental feature of quantum
phenomena.

What is the source of non-objectivity? Operationally, as was
pointed by Bohr \cite{BR}, it is the contribution of a measurement
device to the result of measurement. However, such an operational
explanation does not imply the logical justification of
non-objectivity.

In this paper, we argue that the matter of quantum non-objectivity
is that, on the one hand, the formalism of QM constructed as a
mathematical theory is self-consistent, but, on the other hand,
quantum phenomena as results of experimenter's performances are
not self-consistent. This self-inconsistency is an effect of that
the language of QM differs much from the language of human
performances. The first is the language of a mathematical theory
which uses some Aristotelian and Russellian assumptions (e.g., the
assumption that there are logical atoms). The second language
consists of performative propositions which are self-inconsistent
only from the viewpoint of conventional mathematical theory, but
they satisfy another logic which is non-Aristotelian. Hence, the
representation of quantum reality in linguistic terms may be
different: from a mathematical theory to a logic of performative
propositions. At the level of mathematical theory, we deal with
linguistic terms, satisfying the Aristotelian assumptions. At the
level of logic of experimenter's performances, we deal with
linguistic terms, not satisfying the Aristotelian assumptions.

Thus, we aim to avoid the ``quantum inconsistency'' by applying
modern tools of symbolic logic for studying intelligent behavior
(performances) and we will show that the quantum behavior
satisfies all the basic properties of performances. Logical tools
for studying human behavior were first proposed in the
20th-century language philosophy. Notice that in philosophy of
language since Ludwig Wittgenstein \cite{Wittgenstein}, John
Searle \cite{Searle}, and John Langshaw Austin \cite{Austin} the
ideas of non-objectivity of our everyday reality have actively
developed within the so-called paradigm of \emph{linguistic
solipsism} (cf. with aforementioned views of Bohr, von
Weizs\"acker, Brukner, Zeilinger). According to this paradigm, we
deal just with linguistic reality if we think or act and cannot go
out of language and return to things themselves. Any fact is
seeable and understandable if and only if this is speakable and
the fact can be described in a language \cite{Weinberg}. So, in
any thinking we are limited by our possible speech acts and in any
activity by speech interactions. Language is a part of our
behavior and the way we are interacting with others, for example
by commanding, requesting, pleading, joking, debating, etc.
Philosophers of language distinguish performative propositions
designating and expressing our behavior from informative
propositions denoting facts. While the \emph{informative
propositions} are truth-functions of the elementary propositions
whose meanings are presented by facts, therefore they always have
references in the real world, the \emph{performative propositions}
are non-objective in principle, we cannot find out any real
references for them. They are self-referent and their meanings are
just their utterances \cite{Wittgenstein}, \cite{Austin}. In this
paper, we will show that {\it some quantum statements should be
considered as performative propositions, as well.}

By emphasizing the role of performative propositions in QM, we
cannot avoid a discussion on the role of {\it free will.} We will
show that in QM the problem of free will is involved in
considerations -- how quantum performances can be thought and
treated as appropriate performative propositions for which there
are no real references, because they (as well as performative
propositions about human interactions) have a non-objective
status.

The logical formalism for studying performative propositions was
proposed in \cite{Schumann4}, \cite{Schumann6}. In \emph{Physarum
Chip Project: Growing Computers From Slime Mould}
\cite{Adamatzky1} supported by FP7, we are going to implement this
formalism among others to build up a programmable amorphous
biological computer. In this computer, logic circuits are
presented by programmable behaviors of \emph{Physarum
polycephalum}. Notice that \emph{Physarum polycephalum} is a
one-cell organism that behaves according to different stimuli
called attractants and repellents and can be considered the basic
medium of simple actions that are intelligent in the human meaning
\cite{Adamatzky2}--\cite{Schumann5}. This biological computer has
some properties of quantum computer, in particular we can perform
the double-slit experiment for \emph{Physarum polycephalum} to
show that logical basics of \emph{Physarum} behaviors are the same
as logical basics of quantum behaviors. This means that we face
performativity, non-objectivity, and self-referentiality not only
in human interactions, but also in QM \cite{Schumann3} and in the
behavior of simplest biological organisms (see also \cite{AS1} for
quantum(-like) models of gene expression).

\section{Self-inconsistency of verification of quantum mechanics: the principles of complementarity
and individual-collective duality}

Typically, discussions on self-inconsistency of QM are based on
the principle of complementarity. We briefly present the most
clear analysis of this problem, complementarity and
self-inconsistency of QM, presented by C. Brukner and A. Zelinger
\cite{BZ}. They pointed that N. Bohr \cite{BR} emphasized that
{\small ``How far the [quantum] phenomena transcend the scope of
classical physical explanation, the account of all evidence must
be expressed in classical terms. The argument is simply that by
the word `experiment' we refer to a situation where we can tell
others what we have done and what we have learned and that,
therefore, the account of the experimental arrangement and the
result of observation must be expressed in unambiguous language
with suitable application of the terminology of classical
physics.''} Then, they remarked that rigorously speaking a system
is nothing else, than a construct based on a complete list of
propositions together with their truth values. For a quantum
system, it can happen that the two propositions are mutually
exclusive. This is a specific case of quantum complementarity.
Therefore, in an attempt to describe quantum phenomena we are
unavoidably put in the following situation. On the one hand, the
epistemological structure applied has to be inherited from the
classical physics: the description of a quantum system has to be
represented by the propositions which are used in the description
of a classical system with the Aristotelian semantics, and, on the
other hand, those propositions cannot be assigned to a quantum
system simultaneously. Now, a natural question arises: How to join
these two, seemingly inconsistent, requirements?

In this paper we show that the problem of self-inconsistency of QM
is even deeper, than self-inconsistency implied by the principle
of complementarity. The essence of quantum self-inconsistency can
be better characterized by the principle of the {\it
individual-collective duality} which can be observed in the
\emph{Physarum} behaviour as well (for more details about this
logic see \cite{Schumann4}, \cite{Schumann6}): there are no
logical atoms and something that seems a logical atom (e.g., an
individual behavior) is in fact a family of other sets (e.g., a
collective behavior, see section \ref{logical atom}).

For example, let us define truth-valuations of QM conventionally,
in the way of Aristotle and Russell:
\begin{itemize}
\item (i) the property $E$ is actual (true) in a given state $S$,
whenever a test of $E$ on any physical object $x$ in $S$ would
show that $E(x)$ is true for every $x$ in the state $S$; \item
(ii) the property $E$ is nonactual (false) in a given state $S$
whenever a complementary property $\neg E$ on any physical object
$x$ in $S$ would show that $\neg E(x)$  is true for every $x$ in
the state $S$.
\end{itemize}

Objects $x$ are interpreted as individuals (logical atoms). Assume
that $x$ mean quanta, $E$, $\neg E$ properties, discovered in the
double-slit experiment, with the following meanings: \\

$E$ := ``\emph{the non-detection of the position of $x$ on the
first screen in one of the two slits and the detection of the
position of $x$ on the registration screen corresponding to
the momentum representation with the interference  picture}'';\\

$\neg E$ := ``\emph{the detection of the position of $x$ on the
first screen in one of the two slits or the non-detection of the
position of $x$ on the registration screen corresponding to
the momentum representation with the interference  picture}''.\\

According to the double-slit tests, we face that $ E(x)$ and $\neg
E(x)$ are true for the same state $S$. Obviously, that is
self-inconsistent.

But we {\it can deny our assumptions that $x$ are logical atoms},
i.e., we can assume that $x$ are not exclusive individuals. For
instance, we can put forward the following self-referent
definition of $x$: $\{x\} = \{a,b\}$, i.e., $x$ is both $a$ and
$b$, where $a= (b, (a))$ and $b= (a, (b))$, i.e., $a =
(b(a(b(a(b\ldots)))))$ and $b = (a(b(a(b(a\ldots)))))$ are two
mutually depended infinite streams. Let $ E(a)$ be true, $\neg
E(b)$ be true, $ E(b)$ be false, and $\neg E(a)$ be false. In this
case, $ E(x)$ and $\neg E(x)$ are true for the same state $S$ and
we cannot logically divide $x$ into $a$ and $b$, because $x$ is a
simple object, although it is not an individual. In this paper, we
will show, how we can deal with these strange non-Aristotelian
objects logically.

In our paper, we are limited just by an observational language.
Notice that the language of any physical theory consists of two
different languages: a theoretical language (formal theory with
axioms and inference rules) and an observational language
(semantics for the theoretical language). The first language
contains theoretical terms, which are understood as expressions
that refer to nonobservable entities or properties. The second
language contains observational terms (observables).

A logic for observables is constructed in the observational
language and a logic for theoretical entities in the theoretical
language. In the early 20th century, there was a philosophical
movement of logicism and its followers claimed that it is possible
to construct a general logic for both observables and theoretical
terms. This general logic could be called ``logical physics''. It
is a part of logic, where logical properties of terms and
propositions in relation to space, time, motion, causality, etc.\
are studied \cite{Zinov'ev}. Usually, logical physics has been
presented by logical tools for reducing propositions with
theoretical entities to propositions with observables. In the
accordance with this task, the theoretical terms are understood as
follows: a term $t$ is theoretical if and only if it holds, for
all methods $m$ of determining its extension, that $m$ rests upon
some axioms of some theory $T$ and otherwise it is observational
\cite{Sneed}. For example, in classical mechanics all methods of
determining the force acting upon a particle appeal to some axioms
of classical mechanics (CM), therefore force is a theoretical term
for CM. The entity of spatial distance does not depend upon the
Newtonian axioms. Hence, it is an observable of CM. Thus, the
reduction procedure, which eliminates theoretical terms of an
axiomatic theory by means of observables, is considered a set of
semantic rules for interpreting propositions with theoretical
terms on propositions with observables.

In the way proposed by R. Carnap and C.-O. Hempel, we can reduce
theoretical entities by the following schemata: $c \Rightarrow (h
\Rightarrow e),$ where $h$ is a proposition in theoretical terms
(hypothesis), $c$ and $e$ are propositions in terms of observables
such that $c$ expresses certain observational conditions, which
are satisfied, $e$ presents suitable detecting devices, which then
have to show observable responses.

It was proven that there are ever theoretical terms which cannot
be reduced to observational terms by any logical schemata. This
circumstance of the existence of irreducible theoretical terms
shows the rigorous limits of logical physics and logicism in
physical sciences at all. For example, in QM there are, first, a
formal physical theory formulated in a theoretical language and,
second, its semantics formulated in an observational language
describing quantum experiments. There is also a logical way to
reduce theoretical entities to observables. This way is presented
by quantum logic (QL). Its logical schemas of reduction are as
follows: $$p, q \text{:= `observable $o$ has a value in a Borel
set $\Delta$'}$$ and such propositions are represented by closed
subspaces of a Hilbert space, $\mathcal{H}$. The set of all such
subspaces forms an ortholattice, $\mathcal{L}(\mathcal{H})$, with
$p \leq q$ defined by `$p$ is a subspace of $q$'. The logical
operations of `and', `or', and `not' are modelled respectively by
the operations of meet (infimum), join (supremum) and
orthocomplement on $\mathcal{L}(\mathcal{H})$. The lattice
$\mathcal{L}(\mathcal{H})$ is atomistic, complete, and
orthomodular (non-distributive).

So, as we see, another concept of truth is defined in QL and this
concept is radically different from the classical
(Aristotelian-Russelian) concept of truth \cite{Garola}, because
the QL schemas of reducing theoretical terms are different a lot
from the Carnap and Hempel's classical manner. The main problem of
QL is that even in non-classical means of interpreting QM there
are irreducible theoretical statements. For example, the QM
explanations of the double-slit experiment cannot be directly
interpreted in QL. Nevertheless, {\small ``quantum logics can be
interpreted as a pragmatic language of pragmatically decidable
assertive formulas, which formalize statements about physical
systems that are empirically justified or unjustified in the
framework of QM. According to this interpretation, QL formalizes
properties of the metalinguistic concept of empirical
justification within QM rather than properties of a quantum
concept of truth''} \cite{Garola}, see also Garola et al.
\cite{Garola1},\cite{Garola2} and Rosinger \cite{Rosinger}. The
Garola's pragmatic extension of QL allows him to define
justifications of theoretical statements which cannot be reduced
directly, e.g., within this extension it is possible to justify
the QM explanations of the double-slit experiment. We can remember
that the irreducibility of theoretic entities can imply even
scientific anarchism: {\small ``Science is an essentially anarchic
enterprise: theoretical anarchism is more humanitarian and more
likely to encourage progress than its law-and-order
alternatives''} \cite{Feyerabend}. Therefore, the pragmatic
approach can explicate many presuppositions of quantum physicists
and their way of reasoning as one of the possible ways.

Our approach to QL is different from the conventional QL with
propositions defined on members of $\mathcal{L}(\mathcal{H})$ and
the Garola's pragmatic extension of this QL. First of all, we
would like to follow the pure logicism that has been reanimated by
unconventional computing recently. In unconventional computing, we
appeal to the following schemata of logical reductions: $I
\Rightarrow (h \Rightarrow O),$ where $h$ is a theoretical
proposition, $I$ are inputs of an unconventional computer (quantum
computer, DNA-computer, \emph{Physarum polycephalum} computer,
etc.) and $O$ are outputs of this computer. In these schemata, $h$
is interpreted as a processor of suitable unconventional computer.

Unconventional computing is not so ambitious as physical theories
such as QM. This new approach to computations completely ignores
theoretical entities if they cannot be applied in designing an
appropriate unconventional (abstract or real) processor. Hence, it
deals just with reducible theoretical terms. In our research, we
found out that the behavioral logic constructed on the observables
of \emph{Physarum polycephalum} and parasites of Schistosomatidae
(Trematoda: Digenea) can be directly applied in the double-slit
experiment with quanta. The basic idea of this behavioral logic is
in the individual-collective dualism that there are no logical
atoms in behaviors. Notice that logical theories for
unconventional computing are always constructed in an
observational language. In our opinion, the propagation of photons
has some similarities with an intelligent propagation of
\emph{Physarum polycephalum} \cite{Schumann2}, parasites of
Schistosomatidae (Trematoda: Digenea) \cite{Schumann2a}, and may
other living organisms. Perhaps, we can claim about a new version
of pantheism and idealism that the same patterns of intelligent
behaviors are observed everywhere -- from quanta to one-cell
organisms and human beings.

\section{Self-inconsistency of verification of theoretical viewpoints on photon}

Quantum optics (as a theoretical formalism) is based on the
well-defined and self-consistent notion of photon. In order to
couple the theory with experiments (i.e., to verify theoretical
terms, to reduce them to observables), we need an operational
definition of photon which can be coupled to its theoretical
definition -- as an excitation of quantum electromagnetic field.
Operationally, we can define photon as a click of a photo-detector
(e.g., A. Zelinger, A. Migdall, S. Polyakov, private discussions).
The main point of our discussion is that such a notion is not
self-consistent in the Aristotelian-Russellian meaning (it has no
sense in their semantics). Although nobody did tell about
self-inconsistency of the photon-click definition, the problem is
known (in other terms) and it can be called the problem of the
{\it existence of photon.} In other words, it is the problem of
verifying our theoretical viewpoints on photon. The basic
experiment on the ``existence of photon'' was performed by
Grangier \cite{Grangier1},\cite{Grangier2}, see \cite{Beck} for
reviews on the present experimental situation; see also
\cite{D1}--\cite{DX} for related experimental
studies.\footnote{Formally, the aim of such experiments is to show
an experimental incompatibility of semiclassical optics with
quantum optics. However, from the foundational viewpoint
experimenters really confront the problem of the existence of
photon. The classical wave can be split by beam splitter, but
photon not. Hence, by checking such a splitting one compares the
classical electromagnetic field model with the quantum systems
model.}

The ideal experiment can be described as follows. There is a
single photon source, beam splitter and two detectors, in each
channel of splitter. If ``photons really exist'', i.e., quantum
electromagnetic field cannot be represented as a classical
electromagnetic wave continuously propagating in space-time, then
only one of two detectors has to click. This click can be
identified with the presence of photon in this concrete detector.

We remark that this experiment is a special realization of the two
slit experiment in ``particle context'', i.e., the experiment in
which both slits are open, but two detectors are in work: one
behind each slit.  The claim that only one of these detectors
clicks (for a single photon source) was considered by Bohr as
justification of the {\it principle of complementarity} -- in
combination with the experiment in which both slits are also open,
but without the detectors behind the slits. The later experiment
represents the wave-like interference behavior.  Thus, the
Grangier type experiment on the ``existence of photon'' is of the
fundamental value for quantum foundations. We shall propose a new
interpretation of the experiments of such type.

We point to the well-accepted experimental fact that one can never
expect that the coincidence clicks (i.e., happening simultaneously
in both detectors) will never occur. There are always the
coincidence clicks and there are many such clicks. Therefore it
was decided to count not the absolute number of coincidence
clicks, but the relative number which is given by the coefficient
of {\it second order coherence} $g^{(2)}(0)$; the number of
coincidences divided by the product of numbers of singles (i.e.,
at each of detectors). In principle, one is fine by getting that
$g^{(2)}(0) < 1.$ Such a result was used to reject semiclassical
field theories. However, in real experiments $g^{(2)}(0)$ is still
relatively large (see sections \ref{PEX}, \ref{HER} for details) and {\it the
claim that photon exists, in the operational sense as the click of
a detector, is not justified.}

In such a situation, the operational (and hence experimentally
verifiable) notion of photon cannot be considered as
self-consistent. Any pair of coincidence clicks, for detectors
$D_1$ and $D_2,$ can be interpreted as that two mutually
complementary events, $A_1,$ photon in $D_1,$ and $A_2,$ photon in
$D_2,$ happened simultaneously. However, logically $A_1$ is
negation of $A_2.$ Thus, {\it at the level of the real phenomena,
the theoretical term `photon' of QM is not verified, moreover it
is self-inconsistent on observables.} In our opinion, a possible
explanation is that the observables for the photon notion are
subordinated to performative regularities of some behavioral
entities. And appropriate propositions in observational terms are
not factual, but performative.

Hence, it would be better simply to recognize this fundamental
self-inconsistency and irreducibility of some theoretical terms on
the observables if we appeal to conventional logic and to try to
proceed towards development of a new quantum theory which would
not be based on the conventional logical tools, including
classical QL. The modern development of information and computer
science provides such a possibility. However, in the 1920s
self-consistency of a mathematical theory in the meaning of
classical logic was a fundamental requirement. Therefore, the
self-consistent mathematical theory was created to describe
physical phenomena, even if there is no way to reduce theoretical
terms self-consistently. Heuristically, self-consistency can be
considered as a sign of objectivity, a quantum event is either
firmly true or false. Obviously, then, as Bohr pointed out, this
is only objectivity of observed phenomena within verifications of
a physical theory, i.e., not ``real objectivity'' which was
discussed in the Introduction. Nevertheless, in this situation
heuristically one wants to have some ``elements of reality''. In
our opinion, the self-consistency of the QM formalism is a main
source of the permanent psychological drama in quantum
foundations: reflections towards objectivity (in various forms,
including nonlocal realism which is rather popular nowadays).

We can summarize the discussion of this section as follows: {\it
Some experiments on the photon existence and on the irreducible
deviation of the coefficient of second order coherence from zero
have demonstrated that the operational notion of photon is not
self-consistent on observables. This circumstance suggests us to
construct a new mathematical formalism for quantum phenomena,
which would be based on a logical system permitting behavioral
entities which are performative and self-referent. Usage of the
present mathematical formalism of QM (which is self-consistent)
will permanently induce the illusion of a possibility of objective
interpretation of QM.}

\section{Non-objectivity from the viewpoint of self-inconsistency on observables}
\label{NON}

Let us consider the measurement of photon's polarization. Suppose
that polarization is the objective property of photon. Thus, the
result of the polarization measurement coincides with this
objective property which was predetermined before measurement.
However, the presence of the coincidence clicks and the
corresponding self-inconsistency of the definition of
polarizations up and down, for the setting $\theta$ of the
polarization beam splitter, puts a statistical constraint on this
objectivity. Let us consider representation of the quantum state
$\Psi$ used for measurement by an ensemble of systems which is
denoted by $\Omega.$  For the setting $\theta,$ let us denote the
ensemble of systems (a subensemble of $\Omega)$  producing the
coincidence clicks by the symbol $\Omega_\theta.$ Hence, the
self-consistent definition of the property of polarization in the
direction $\theta$ is possible only on the subensemble
$\bar{\Omega}_\theta = \Omega \setminus \Omega_\theta =\{\omega
\in \Omega: \omega \not\in \Omega_\theta\},$ the complement to
$\Omega_\theta.$ Therefore, the vector of polarization can be
objectively (and consistently) defined only on the subensemble
$\tilde{\Omega} \equiv \Omega \setminus \bigcup_\theta
\Omega_\theta = \bigcap_\theta \bar{\Omega}_\theta.$ Of course,
 for each fixed $\theta,$ the probability
of the coincidence clicks is very small, $P(\Omega_\theta)
=\epsilon <<1.$\footnote{We emphasize that in any real
experimental setup, although this probability is small, it has
nonzero low bound: $P(\Omega_\theta) \geq \epsilon_{\rm{min}} >0,$
see sections \ref{PEX}, \ref{HER}.} However, the probability of
the union $\bigcup_\theta \Omega_\theta$ can be close to one. (In
the complementary terms, although $P(\bar{\Omega}_\theta )\approx
1,$ it can happen that the probability $P(\tilde{\Omega})=
P(\bigcap_\theta \bar{\Omega}_\theta) \approx 0.$) Thus, the
self-inconsistency of polarization's observable in the form of the
presence of coincidence clicks can restrict the possibility of the
objective definition of polarization to a very small subensemble
of systems prepared in the state  $\Psi.$

The logical possibility that the ``objectification subensemble''
$\tilde{\Omega},$ can have approximately zero
probability\footnote{To be more rigorous, one has to speak about
the probability to draw a system from the objectification
subensemble $\tilde{\Omega}$.}, makes the project of
objectification of quantum observables questionable. In the light
of the previous consideration, the appearance of non-objectivity
in QM is not so surprising. Hence, if one has doubts in quantum
non-objectivity, he has to find strong reasons for this.

However, in principle, to the question of objectification we need
not proceed under the aforementioned assumption that elements of
$\bigcup_\theta \Omega_\theta$ form a representative sample (or in
complementary terms that  the objectification subensemble
$\tilde{\Omega}$ is a non-representative sample). In order to
destroy objectification, it is sufficient to use the
experimentally justified assumption that the probability of each
$\Omega_\theta$ is sufficiently far from zero: $P(\Omega_\theta)
=\epsilon,$ where one can take $\epsilon \approx 0,03$ for sources
producing photons on demand, see section \ref{HER}. In such a
situation, we cannot proceed with objective polarization, simply
because we do not know whether for the coming trial the result
will self-consistent or not. We may get a single click in one of
channels, but we also may get coincidence clicks.

We may summarize the results of our analysis of the inter-relation
of (non-)objectivity of quantum observables and their
self-(in)consistency from the classical viewpoint in the follofing
way: {\it The self-inconsistency of reducing theoretical entities
of QM to {\bf performative descriptions} of quantum observables
makes really impossible the objectification of QM observables.
(The procedure of objectification with some probability definitely
contradicts to the standard views on objective reality.)}

\section{Self-inconsistency contra elements of reality of Einstein, Podolsky, and Rosen}

The Einstein, Podolsky, and Rosen (EPR) argument based on the
consideration of elements of reality corresponding to quantum
observables measured for some specially prepared states \cite{EPR}
(which are nowadays known as entangled states) is one of the
strongest motivations for attempts of the objective interpretation
of quantum observables. For example, in his considerations leading
to Bell's inequality, Bell pointed that there is the strong reason
to consider quantum observables as objective, precisely because of
the EPR-argument \cite{B}. The EPR-derivation of the possibility
to assign to quantum  systems in some states the objective values
of two incompatible quantum observables was criticized in
\cite{Luders} from the viewpoint of usage of L\"uders projection
postulate in the case of observables with degenerate spectrum,
instead of the von Neumann original postulate. Now, we try to
destroy the EPR-argument by using self-inconsistency argument of
performances on observables.

Again, as in section \ref{NON}, we can decrease the probability of
objectification in the EPR-experiment by considering families of
incompatible quantum observables. However, as was pointed in the
previous section, for our purpose we need not proceed in such a
way. Even for the fixed observable, we cannot predict, whether the
result of the coming trial would permit the objectification or
not. (Here we discuss the quantum optics version of the
EPR-experiment, in which the projections of the photon
polarization on different axes play the role of the original
EPR-observables, position and momentum.)

The latter argument shows that the essence of the objectification
problem is not in the presence of incompatible quantum
observables.

Now in the light of our approach, we can remember the Bohr's reply
to the EPR-argument \cite{BR_REP}. Bohr stressed that even for one
fixed setting one is not able to assign the element of reality to
the first component of a compound system on the basis of the
result of measurement on the second component. Thus, he also
pointed that the problem arose already in the case of a single
observable. His conclusion matches very well with our conclusion
(although Bohr did not paid attention to self-inconsistency of
theoretical terms on quantum observables).

\section{Free will (performativity) against self-incon\-sistency}

Various ``technicalities'' (see, e.g., the discussion below) play
important roles in quantum experiments. These technicalities are
not presented in the mathematical formalism of QM. Therefore, the
real outputs of experiments deviate from the theoretical
predictions based on straightforward mathematical computations.
Taking these technicalities into account is a difficult problem.
(In fact, it can be treated as a part of the quantum measurement
problem.) Nevertheless, we can pay attention that some basic
elements of these technical issues of the design of concrete
experiments can be considered self-referent performative
propositions. We illustrate this situation by consideration of
quantum optics experiments.

All quantum optics measurements are fundamentally based on the
proper choice of the {\it discrimination threshold.} It is a kind
of performance that allows us to understand quantum phenomena. For
our argument, it is very important to remark that the setting of
the sufficiently high discrimination threshold is an important
part of experiments on ``photon existence'', otherwise the
$g^{(2)}(0)$ coefficient would be too large; see the Grangier's
PhD-thesis \cite{Grangier1}: ``[...] In this configuration, the
threshold has a double role of acquisition of timing information
and of selection of the pulses (the too weak pulses are not taken
into account). [...]
We have in the present experiment chosen a {\bf rather high
threshold}, which amount to give the priority of the stability of
the counting rates and the reproducibility of the results, rather
than to the global detection efficiencies.'' (We stressed with
bold the important fact that Grangier proceeded with rather high
threshold.) Thus, in our terms the selection of the discrimination
threshold plays the fundamental role in minimization of
self-inconsistency of reductions to quantum observables.

Evidently, the standard interpretation of this choice of the
discrimination threshold is that this is the noise minimization
procedure. However, if one uses the operational definition of
photon as a click of a detector, i.e., if one studies the real
quantum phenomena and not just theorizing, then there is a problem
of separation of ``noisy photons'' from ``real photons'', since
both types are just clicks of detectors.

By putting the discrimination threshold, we insert a subjective
element in all quantum optics measurement schemes. This insertion
is based on our {\it free will} -- to minimize self-inconsistency
of QM (more concretely, self-inconsistency of the operational
definition of photon as detector's click) by appropriate
performances. This is a complex psychological play. First, the
scientists created a self-consistent mathematical representation
of quantum phenomena. Then, they confronted the problem of
coupling of theory with experiments. This is really a shadowed
area of QM. One would not find so much material on coupling of
theoretical entities of QM with {\it real experiments.} Typically,
one is completely fine by repeating the Bohr's statement that in
some experimental contexts photons exhibit particle features. What
is the experimental reality to be a particle, for photon? It seems
that this important problem is practically ignored in theoretical
studies on quantum foundations. However, experimenters have to
solve this problem in everyday life. They do not discuss it in the
papers presenting results of experiments, and majority simply
ignores it. However, foundation-thinking experimenters understand
the importance of this problem and each of them solves it for
himself; and surprisingly, the solution is the same:
experimentally, photon is nothing else than the detector's click.
However, by our interpretation experiments (of Grangier's type)
showed that this operational definition of photon is not
self-consistent within classical QL.  Again by our interpretation,
an experimenter minimizes self-inconsistency with the aid of his
free will.

Of course, if one considers free will as just a mental illusion
and uses the picture of the totally deterministic universe, see K.
Svozil \cite{Svozil} for the detailed analysis of such a position,
also cf. G.`t Hooft \cite{TH2}, then nature by itself minimizes
self-inconsistency in the quantum phenomena. However, our analysis
showed that even nature would not be able beat self-inconsistency
completely: it has to respect the statistical constraint based on
the presence of the coincidence clicks.

We also point to usage of another subjective element in minimizing
self-inconsistency of coupling of quantum observables (as elements
of the mathematical model of quantum mechanics) with real
experiment. This is selection of the time window for
identification of the clicks in the two channels of a beam
splitter as the coincidence clicks. The coefficient $g^{(2)}(0)$
fundamentally depends on this time window. Here again free will
plays an important role. This is a performance type element of QM.
The size of the time window is determined subjectively aiming to
reproduce predictions of the QM mathematical model. We remark that
the role of a proper selection of time window was discussed in
very details in connection with the Bell type tests, see
\cite{Hess}--\cite{Gill}, \cite{Luders}. This is well known
coincidence time loophole for these test. In this paper, we point
out that the same problem arises not only in experiments with
entangled photons, but even in single photon experiments.

In general without subjective determination of ``technicalities''
such as  thresholds and time widows, an experimenter is not able
to approach even approximately matching with the QM theoretical
formalism. Notice that these ``technicalities'' are not elements
of mathematical formalism of QM.

We could summarize the discussion on free will, performativity,
and self-(in)consistency of QM as follows: {\it Experimenter's
free will plays the crucial role in the improvement of
self-inconsistency of quantum observables. Selections of proper
values of various ``experimental technicalities'' can be
interpreted as attempting to lower self-inconsistency of QM
presented in usage of performance type statements in establishing
coupling between theory and experiment. Not doing it
intentionally, experimenters construct performative propositions
for which there is another, non-Aristotelian logic.}

\section{Experiments on ``photon existence''}
\label{PEX}

It is well known that photomultipliers and
silicon-avalanche-photodiodes  have low efficiency: an essential
part of the ensemble $\Omega$ of quantum systems representing some
quantum state, say $\Psi,$ disappears without any click. The
presence of the ``no-detection'' event  also contributes to
self-inconsistency of theoretical terms on quantum physical
phenomena. For some setting $\theta,$ the two events, $A_1$ --
``polarization up'' and $A_2$ -- ``polarization down'', are
considered as complementary and appearance of the third even,
$A_3$ -- ``no detection'', destroys self-consistency on quantum
observations, even in the absence of coincidence clicks.
Therefore, from the very beginning we have considered the
experiments on ``photon existence'', on estimation of the
coefficient of second order coherence, which were done with
detectors of low efficiency as just dimming the problem of
self-(in)consistency of quantum entities due to the presence of
the coincidence clicks.

We are interested in experiments of the aforementioned type for
detectors of very high efficiency, for TES-detectors.
Theoretically they have 100\% efficiency.

However, the main problem is even not in the detectors
inefficiency. The main problem is that in reality {\it there are
no pure single photon sources:}

``An ideal single-photon source would be
one for which: a single photon can be emitted at any arbitrary
time defined by the user (i.e., the source is deterministic, or
``on-demand''), the probability of emitting a single photon is
100\%, the probability of multiple-photon emission is 0\%,
subsequent emitted photons are indistinguishable, and the
repetition rate is arbitrarily fast (limited only by the temporal
duration of the single-photon pulses, perhaps),'' see \cite{39}.

Although in literature one may read about single photon sources,
this is merely a terminological trick. There is a fundamental
limit of ``single photonity'': if the temperature is higher than
zero (Kelvin), then in principle a black body which is always
present in the experimental setup can radiate a photon in the
prepared mode. In optics such a probability (for the room
temperature) is very small, but it is, nevertheless, nonzero. To
the most part, getting a real on-demand source that would produce
an appreciable amount of photons is hard. On-demand sources that
are readily available suffer from low single photon purity, with
$g^{(2)}(0)=0.07.$ Some heroic efforts have led to lower
$g^{(2)}(0)$, but these sources are too dim, hard to align and
keep aligned, etc.

Nowadays, it is quite common to refer as a ``single photon
source'' to a source such that $g^{(2)}(0) < 0.5.$ Such an
approach, namely, usage of the coefficient of second order
coherence to determine whether a source is of the single photon
type and then, for such sources, to measure the same coefficient
to establish the operational notion of photon, is definitely based
on the argument of the {\it circular type}  -- this is a
consequence of the irreducible self-inconsistency of the quantum
theoretic terms, at least of quantum optics.

We can, finally, say that: {\it One has to be well aware that the
expression ``a single photon source'' is simply jargon used by
experimenters. Unfortunately, by the theoretical part of the
quantum community this expression was taken too straightforward.
The usage of the coefficient of second order coherence for the
operational definition of a single photon source (although
acceptable operationally) is totally unacceptable foundationally.
The experimental groups working on problems related to foundations
of quantum optics have to put new efforts to create much better
approximations to single photon sources. Finally, clean
experiments to estimate the coefficient of second order
interference with such on-demand sources and TES-detectors have to
be performed. Such experiments are difficult to perform. And one
of the psychological problems preventing to put essential efforts
to such experimental studies is that there is a very common
opinion that the question about the ``existence of photon'' has
already been totally clarified. This is the wrong viewpoint. There
was done only the easiest part of experimental studies, by using
bad sources and bad detectors, which can be considered as only a
preparatory stage for future real foundational studies in
experimental quantum optics.}

\section{Non-objectivity from the viewpoint of performativity}
\label{logical atom}

Let us recall that the Kolmogorov's main assumption in probability
theory is that there exists a set partition into disjoint subsets
and, respectively, the probability measure defined on the given
set is calculated as the addition of appropriate probabilities
defined on subsets. However, we have just exemplified in the
previous sections that there are observables, where the additivity
for probabilities is falsified if we deal with behaviors of
quanta, living organisms, etc.

The \textit{intuition of objectivity} that has been felt by the
majority of physicists since the Ancient times till now was first
formulated by Aristotle. According to him, there is
`\textit{hypokeimenon}' as substratum of any predicates.
\textit{Hypokeimenon} is a family of singular events or singular
facts (`atoms' in the first meaning proposed by Democritus). For
quantum physicists, \textit{hypokeimenon} is given by smallest
particles and all the world is described by predicates in relation
to these particles, like that: `the quantum has the property $A$',
`the complex $B$ of quanta has properties $A_{x} $, $A_{y} $,
$A_{z} $, \dots, which explore physical phenomena $x$, $y$, $z$,
\dots, respectively', \dots, etc. By Kolmogorov, probabilities
should be involved in our reasoning just on particles (singular
events) or their Boolean compositions. However, the double-slit
experiment means that photons cannot be considered the
Aristotelian \textit{hypokeimenon} and there are no singular
events at all.

According to Aristotle, \textit{hypokeimenon}, the `first
subject', underlying things $a$, $b$, $c$, \dots, present an
objective reality. Every underlying thing possesses unique
properties. It means that $a$, $b$, $c$, \dots  are atoms of our
database. There is nothing less than them. Due to properties, we
can group atoms within different classes $P$, $Q$, $R$, \dots  The
more general property of thing, the more extensive class to which
it belongs by this property.

Hence, the idea of \textit{hypokeimenon}, the underlying things,
allowed Aristotle to build up formal databases as well-founded
trees of data (i.e., these trees are finite and without cycles or
loops). He started with underlying things as primary descendants
of trees in constructing ontological (syllogistic) databases. Let
us notice that, by Aristotle, different sciences have different
syllogistic databases, because they use different means for
obtaining predicates for \textit{hypokeimenon}. The quantum
physics follows this Aristotelian understanding of objectivity and
differs from the Ancient physics only by different ways of
creating predicates for underlying things which are understood now
as smallest particles.

Syllogistic trees contain genus-species relations among items. We
know that in genus-species relations we can consider a branch (a
relation between a genus and species) as implication, where the
top of branch (genus) is regarded as consequent of implication and
the bottom of branch (species) as antecedent of implication. Then
for each node of the genera-species tree, we may define an
intension as all reachable genera (all higher nodes) and an extent
as all reachable species (all lower nodes). It is known that the
greater extent, the smaller intension and the greater intension,
the smaller extent.

Thus, the first logical database was invented by Aristotle. It is
designed in his syllogistics. He suggested using this database as
a logical frame for different sciences. Therefore if we claim that
a science is a database constructed on the basis of empirical
observations by applying logical inference rules, then we can
claim that the history of exact science has started since
Aristotle.

Let us assume that such a database is closed under all logical
operations. Then in this database the following relations take
place. Let \textit{P} be a property. Then there is also a property
non-\textit{P}. Further let \textit{P} be more extensive than
\textit{Q} (i.e. an appropriate class \textit{P} is more
extensive, than \textit{Q}). Then we obtain the following
relations (see figure \ref{fig:1}):

\begin{figure}
\centering
\includegraphics[width=2.25in]{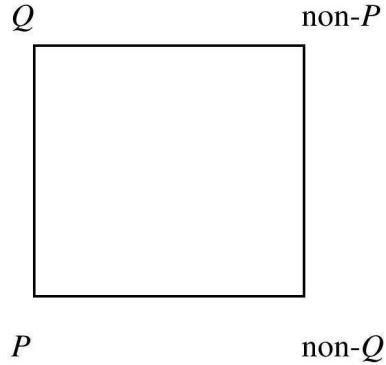} \caption{The Aristotelian square of oppositions.}
\label{fig:1}\end{figure}

\begin{itemize}
    \item $Q$ and non-$P$ are properties
which can be together false, but they cannot be together true in
relation to any atom of our database;
    \item $P$ and non-$Q$ are properties
which can be together true, but cannot be together false in
relation to any atom of our database;
    \item $P$ and non-$P$ are properties which cannot be
together true and cannot be together false in relation to any atom
of our database;
    \item $Q$ and non-$Q$ are properties
which cannot be together true and cannot be together false in
relation to any atom of our database;
    \item if $Q$ is true in relation to some
atoms of our database, then also \textit{P} is true in relation to
the same atoms of our database;
    \item if $P$ is not true
in relation to some atoms of our database, then also $Q$ is not
true in relation to the same atoms of our database.
\end{itemize}

Thus, in the Aristotelian database we deal with the Boolean
algebra due to the assumption of existence of logical atoms
(singular events). Since Aristotle the objectivity has been
understood as a possibility to construct databases when there is
`subject', the underlying things, the family of atoms grouped in
classes, so that these classes are closed under all logical
operations. In different sciences we choose different properties
of atoms and as a consequence we group atoms differently. Such an
intuition of objectivity holds in quantum physics till now.

Notably, classical mechanics (CM) can be readily presented as a
semantics for the Aristotelian logic closed over logical
superpositions of syllogistic propositions of the following kind:
``All $S$ are $P$'', ``Some $S$ are $P$'', ``No $S$ are $P$'',
``Some $S$ are not $P$''. The point is that in CM, first, we have
Aristotelian atoms or individuals defined as particles, second, in
CM the state of a system $\mathcal{S}$ consisting of $N$ particles
is defined by giving the $3N$ position coordinates and the $3N$
momentum coordinates. Hence, according to CM, any state of
$\mathcal{S}$ is fully determined by three values for position and
three values for momentum of all particles of $\mathcal{S}$. This
two circumstances allow us to define a verification of syllogistic
propositions as follows:
\begin{description}
    \item[$S \mathbf{a} P$:] ``All particles of $\mathcal{S}$ with positions $S$ have momentums $P$'' means that the
    system $\mathcal{S}$ is not empty (i.e., it contains some
    particles) and for all particles of $\mathcal{S}$ if we know
    their position $S$, then we know their momentum $P$;
    \item[$S \mathbf{i} P$:] ``Some particles of $\mathcal{S}$ with positions $S$ have momentums $P$'' means that for
    some particles of the system $\mathcal{S}$ we know
    their position $S$ and we know their momentum $P$;
    \item[$S \mathbf{e} P$:] ``No particles of $\mathcal{S}$ with positions $S$ have momentums $P$'' means that for
    all particles of the system $\mathcal{S}$ we do not know
    their position $S$ or we do not know their momentum $P$;
    \item[$S \mathbf{o} P$:] ``Some particles of $\mathcal{S}$ with positions $S$ do not have momentums $P$'' means that all particles do not belong to the
    system $\mathcal{S}$ or there are particles of $\mathcal{S}$ such that we know
    their position $S$ and we do not know their momentum $P$.
\end{description}
Formally:
\begin {equation}S \mathbf{a} P := (\exists A (A \varepsilon S) \wedge \forall
A (A \varepsilon S \Rightarrow A \varepsilon P));\label{for:1}\end
{equation}
\begin {equation}S \mathbf{i} P := \exists A (A \varepsilon S \wedge A
\varepsilon P);\label{for:2}\end {equation}
\begin {equation}S \mathbf{e} P := \neg(S \mathbf{i} P);\label{for:3}\end {equation}
\begin {equation}S \mathbf{o} P := \neg(S \mathbf{a} P).\label{for:4}\end {equation}
All other propositions of Aristotelian logic are defined thus: (i)
each syllogistic proposition defined in
(\ref{for:1})--(\ref{for:4}) is a proposition, (ii) if $X,Y$ are
propositions, then $\neg X$, $\neg Y$, $X \star Y$, where $\star
\in \{\vee, \wedge, \Rightarrow\}$, are propositions, too. Now,
the Aristotelian logic can describe properties of our knowledge on
systems $\mathcal{S}$ of CM.

Nevertheless, we can assume reality without objectivity, i.e.,
without atoms of databases. In modern logic universes in which
there are no atoms are studied as well. However, in modern
sciences the intuition that logical atoms exist has been used till
now, and Aristotle's reasoning has been intuitively applied.

\begin{figure}
\centering
\includegraphics[width=2.25in]{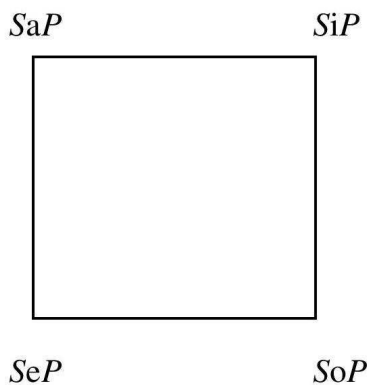} \caption{The unconventional square of oppositions.}\label{fig:2}
\end{figure}

Notice that any context-based reasoning can be realized only in a
universe without atoms. For instance, let us consider the
following two propositions from the Bible: `bestow that money for
sheep' and `bestow for whatsoever thy soul desireth'
(\textit{Deut}. 14:26). Syntactically, if we assume the existence
of logical atoms, `bestow for whatsoever thy soul desireth' is a
universal affirmative proposition ($S \mathbf{a} P$) and `bestow
that money for sheep' is a particular affirmative proposition ($S
\mathbf{i} P$), i.e., the first is more general, than the second.
However, for example, I do not desire sheep and I do not know
people who desire it. Perhaps, such people exist, but I do not
know. Then we cannot plot the classical square (figure
\ref{fig:1}), because `bestow that money for sheep' is not
included into `bestow for whatsoever thy soul desireth', e.g.,
maybe my soul does not desire sheep, but desire many other things.
The matter is that `thy soul desireth' is a performative
proposition and it has different meanings at different situations
(there are no atoms for that proposition). This means that in this
Biblical example the implication $S \mathbf{a} P\Rightarrow S
\mathbf{i} P$ is false in general case. Thence, we could assume
another semantics, where $S \mathbf{a} P$ and $S \mathbf{i} P$ are
different viewpoints of the same level. Therefore, at one and the
same situation of utterance both statements (`bestow for
whatsoever thy soul desireth', $S \mathbf{a} P$, and `bestow that
money for sheep', $S \mathbf{i} P$) may be simultaneously false,
but cannot be simultaneously true. In this way we obtain the
unconventional square of opposition (figure \ref{fig:2}).

The same situation takes place for photons and other quanta. The
logical square of figure \ref{fig:1} does not hold for them,
because logical atoms do not exist for all situations (for the
double-slit experiment when both slits are open). Let us consider
the propositions `the photon can be detected passing through all
slits' ($S \mathbf{a} P$), `the photon cannot be detected passing
through the slits' ($S \mathbf{e} P$), `the photon can be detected
passing through some slits (probably through the one slit)' ($S
\mathbf{i} P$), `the photon cannot be detected passing through
some slits (probably through the one slit)' ($S \mathbf{o} P$). If
it is possible to say that the following implication is valid: if
`the photon can be detected passing through all slits', then `the
photon can be detected passing through some slits (probably
through the one slit)'? In other words, if the photon is a wave
and described in a proposition in theoretical terms of `wave',
then the same photon is a particle and described in another
proposition in theoretical terms of `particle'? Rather there
should be the disjunction: $S \mathbf{a} P$ or $S \mathbf{i} P$.
But not implication. Indeed, if $S \mathbf{a} P$ is true, then the
photon is not a particle in common meaning, and if $S \mathbf{i}
P$ is true, then maybe the photon is a particle in common meaning.

In QM, it is impossible to define logical atoms. Indeed, if there
were logical atoms in QM, our propositions would not be
performative and would not depend upon the context of quantum
experiments. It means that the reality exists, but it does not
correspond to the Aristotelian intuition of objectivity when the
underlying things exist, i.e., any behavior is reduced to an
individual behavior. Instead of this intuition, we can propose
another intuition, the non-well-funded objectivity
\cite{Schumann3}, when there are no logical atoms, i.e., there is
no `first subject' in the Aristotelian meaning. We can add that in
our picture of the world there are no things themselves
(`\emph{Dinge an sich}' in the Kantian meaning), nothing, only
behavioral complexes described by self-referent performative
propositions. It is a kind of idealism proposed in the linguistic
solipsism of Wittgenstein and accepted by us. A similar idealistic
approach was proposed by E. Husserl. He states that phenomena are
nothing more than our consciousness, and pure phenomenology is the
science of pure consciousness: {\small ``Natural objects, for
example, must be experienced before any theorizing about them can
occur. Experiencing is consciousness that intuits something and
values it to be actual; experiencing is intrinsically
characterized as consciousness of the natural object in question
and of it as the original: there is consciousness of the original
as being there `in person' ''} \cite{Husserl}. Thus, {\small ``the
concept `phenomenon' carries over, furthermore, to the changing
modes of being conscious of something -- for example, the clear
and the obscure, evident and blind modes in which one and the same
relation or connection, one and the same state of affairs, one and
the same logical coherency, etc., can be given to consciousness''}
\cite{Husserl}. Notice that Gestalt psychology is based on these
ideas of Husserl.

Thus, in non-well-founded objectivity there are no logical atoms.
What does it mean? We have properties (classes) $P$, $Q$, $R$,
\dots Some of these classes have non-empty intersections, and some
others do not. Logical atoms are classes which cannot be
intersected at all, they are singletons. Their intersection is
always empty. This fact can be considered the definition of
logical atom. Their logical combination (disjunction, conjunction,
complement)  gives any class $P$, $Q$, $R$, \dots Accordingly, the
universe in which there are no logical atoms is universe in which
the intersection of classes is not empty. Different combinations
of these intersections give different contexts of performative
propositions. For such a universe instead of the Aristotelian
square of opposition (figure \ref{fig:1}), another square takes
place (figure \ref{fig:2}), e.g.,
\begin{itemize}
    \item \emph{In the double-slit experiment with photons}:

    `The photon can be detected in all slits'
($S \mathbf{a} P$), `the photon can be detected in no slits' ($S
\mathbf{e} P$), `the photon can be detected in some slits (in QM,
it means, just in one)' ($S \mathbf{i} P$), `the photon cannot be
detected in some slits (just in one)' ($S \mathbf{o} P$).

    \item \textit{Formally}:

\begin{enumerate}
    \item $S \mathbf{a} P$ and
$S \mathbf{i} P$ are properties which can be together unjustified,
but they cannot be together justified in relation to any
performative situation of our database;
    \item $S \mathbf{e} P$ and $S \mathbf{o} P$ are properties which can be together justified, but cannot be together
unjustified in relation to any performative situation of our
database;
    \item $S\mathbf{a}P$ and
$S \mathbf{o} P$ are properties which cannot be together justified
and cannot be together unjustified in relation to any performative
situation of our database;    \item $S \mathbf{e} P$ and $S
\mathbf{i} P$ are properties which cannot be together justified
and cannot be together unjustified in relation to any performative
situation of our database; 
\item if $S \mathbf{a} P$
(respectively, $S \mathbf{i} P$) is justified in relation to some
performative situations of our database, then also $S \mathbf{e}
P$ (respectively, $S \mathbf{o} P$) is justified in relation to
the same situations of our database\footnote{We should remark that in logic, the implication does not
mean a causal relation or deep semantic relationship between
antecedent and consequent. For example, the sentence ``2+2=4''
implies that ``We are born in the USSR'', because both sentences are true.
In case of syllogistic, we group quantified propositions into some
classes according to their truth-conditions. So, this relationship
between antecedent and consequent just formally follows from our
formal definitions of $S \mathbf{a} P$, $S \mathbf{e} P$, $S
\mathbf{i} P$, $S \mathbf{o} P$, see (\ref{for:5})--(\ref{for:8})
and their interpretations on quantum observables below. In an
informal interpretation, this relationship between antecedent and
consequent means that the class of events of non-detecting in both
slits is largest. The class of detecting in one slit or in both
slits is smallest. It directly follows from our formal definitions
below.};
\item if $S \mathbf{e} P$
(respectively, $S \mathbf{o} P$) is not justified in relation to
some performative situations of our database, then also $S
\mathbf{a} P$ (respectively, $S \mathbf{i} P$) is not justified in
relation to the same situations of our database.\end{enumerate}
\end{itemize}

Notice that in our version of QL, performative propositions
expressing observables are not true or false in a conventional
meaning of Russellian-Tarskian semantics, but they are justified
or unjustified. Really, they have a pragmatic rather than a
semantic interpretation. In the Austian semantics (also called the
situation semantics), they are evaluated as successful or
unsuccessful in the given situation of utterances. We use some its
versions in our logic of self-referent performative propositions.

So, instead of atoms in the quantum universe we deal with
performative situations, i.e., with different intersections of
classes (properties) in a collective behavior. The logical theory
of performative propositions was proposed in \cite{Schumann4},
\cite{Schumann6}. In this theory we obtain non-well-founded
syllogistic trees for which there cannot be underlying things
(\textit{hypokeimenon}). Thus, there is no objectivity in
classical meaning. Indeed, we can always define intersections $A\&
B$ for some situations $A$ and $B$ such that $A\& B$ is an infimum
of $A$ and $B$. Therefore there are no atoms which can be used for
building trees-molecules as their superpositions. Instead of
underlying things, we suppose situations that can always be
intersected.

The Aristotelian logic with syllogistic propositions defined in
(\ref{for:1})--(\ref{for:4}) is self-inconsistent on quantum
observables, although CM plays the role of semantics for this
logic, as we said. Nevertheless, we can offer a non-Aristotelian
system without logical atoms, where syllogistic propositions have
the following meanings:
\begin{description}
    \item[$S \mathbf{a} P$:] ``All quanta of $\mathcal{S}$ with positions $S$ have momentums $P$'' means that the
    system $\mathcal{S}$ is not empty (i.e., it contains some
    quanta) and for all experiments with quanta of $\mathcal{S}$
    their position $S$ is absolutely uncertain for all possibilities and we know their momentum $P$; \emph{in the
    double-slit experiment}: the
    system $\mathcal{S}$ is not empty and for all experiments with quanta of $\mathcal{S}$
    these quanta pass through both slits ($S$) and their momentum has an interference picture
    ($P$);
    \item[$S \mathbf{i} P$:] ``Some quanta of $\mathcal{S}$ with positions $S$ have momentums $P$'' means that for
    all quanta of the system $\mathcal{S}$ we know
    their position $S$ and we do not know their momentum $P$; \emph{in the
    double-slit experiment}: for all experiments with quanta of $\mathcal{S}$
    these quanta pass through one slit ($S$) and their momentum does not have an interference picture
    ($P$);
    \item[$S \mathbf{e} P$:] ``No quanta of $\mathcal{S}$ with positions $S$ have momentums $P$'' is justified iff ``Some quanta of $\mathcal{S}$ with positions $S$ have momentums $P$'' is not justified;
    \item[$S \mathbf{o} P$:] ``Some quanta of $\mathcal{S}$ with positions $S$ do not have momentums $P$'' is justified iff ``All quanta of $\mathcal{S}$ with positions $S$ have momentums $P$'' is not justified.
\end{description}
Formally:
\begin {equation}S \mathbf{a} P := (\exists A (A \varepsilon S) \wedge \forall
A (A \varepsilon S \wedge A \varepsilon P));\label{for:5}\end
{equation}
\begin {equation}S \mathbf{i} P := \forall A (\neg (A \varepsilon S) \wedge \neg (A
\varepsilon P));\label{for:6}\end {equation}
\begin {equation}S \mathbf{e} P := \neg(S \mathbf{i} P);\label{for:7}\end {equation}
\begin {equation}S \mathbf{o} P := \neg(S \mathbf{a} P).\label{for:8}\end {equation}
All other propositions of non-Aristotelian quantum logic are
defined as follows: (i) each syllogistic proposition defined in
(\ref{for:5})--(\ref{for:8}) is a proposition, (ii) if $X,Y$ are
propositions, then $\neg X$, $\neg Y$, $X \star Y$, where $\star
\in \{\vee, \wedge, \Rightarrow\}$, are propositions, also. This
non-Aristotelian logic is a very simple version of QL without
logical atoms. All its propositions are performative and depend
upon contexts.

Hence, we could claim that: \emph{From the viewpoint of
performativity and logical theories studying performative
propositions, there is no objective reality in the classical
(Aristotelian) meaning. In QM, scientists try to appeal to the
objective reality with logical atoms of quantum systems, which
causes self-inconsistencies. Therefore, the only outcome is in
appealing to non-well-founded reality \cite{Schumann3} and
performative propositions in QM. Self-inconsistency occurs only in
cases of applying classical logic and classical semantics. In our
logic, there are no contradictions. The same situation is in the
so-called paraconsistent logics, where there are contradictions as
new truth-values. Self-inconsistency is just in that we avoid
logical atoms and even contrary statements in the classical logic
may have non-empty intersections in our new semantics. Thus, the
term self-inconsistency only concerns logical properties of our
version of QL and not phenomena. Evidently, phenomena themselves
simply occur and cannot be self-inconsistent.}

\section{Existence of photon from the viewpoint of heralded photons}
\label{HER}

Due to applications of ideas of performativity in QM to the
problems of single photon on demand source (low coefficient of
second order coherence, high brightness, experimental feasibility,
etc.), it is possible to propose a workable solution consisted in
{\it heralded photon sources}, i.e., sources based on parametric
down-conversion. These sources produce low levels of classical
light in each of the two conjugated modes, but have a property
that {\it photons are created in pairs}: one per conjugated mode,
therefore detecting one photon in one mode means that a photon in
the other was created with 100\% certainty. In good experiments it
is possible to collect up to 70\% of these heralded photons. We
therefore write that conditional
$P(\rm{detection}_1|\rm{emission}_2)=0.7,$ where 1,2 are the mode
numbers. Similarly, $g^{(2)}_{\rm{conditional}}(0)=0.01.$

One can think of such a source as a source of pulsed ``single
photons'', where one learns about the presence of a good single
photon (with $P_{\rm{register}}=0.7$ per pulse) by seeing a click
in the other mode. Most of experimenters use these sources and
call them single photon sources. This substitution may be
justified from the operational viewpoint, but it makes a big
difference from the foundational viewpoint.

From the viewpoint of self-(in)consistency analysis of quantum
physical entities, by using heralded photon sources, the
experimenters try to minimize self-inconsistency by considering
conditioned events. This is the crucial departure from the
original event structure of the quantum formalism which is based
on the assumption of the existence of individual quantum systems
and uses the event algebra (in fact, Boolean) to describe
measurements of a single observable on such systems. Moreover,
30\% of unused pairs also destroys consistency of the event
structure of the yes-no experiments, i.e., we can claim that there
is no Aristotelian objectivity with logical atoms.

Finally, we emphasize that, although the value
$g^{(2)}_{\rm{conditional}}=0.01$ is relatively small, the number
of coincidence clicks is still non-negligible. Hence, we can
repeat considerations of section 6 and derive non-objectivity of
photonic observables from self-inconsistency of theoretic entities
on observables. Here, the probability of coincidence also has
nonzero low bound which, of course, depends on the experimental
setup. It depends on so many ``technicalities'' that its
calculation is really a nontrivial task.  First and foremost, one
needs to know the number of pairs generated per second. One can
expect that $g^{(2)}(0)$ is higher for brighter states, and lower
for dimmer states. Losses in both heralding and experimental
channel are also important as well as properties of
down-conversion. If one operates with a single mode source, the
noise photons would have thermal, and not Poisson statistics. If
one had a low mode number, statistics would be a finite sum of
thermal states, and if one had infinite number of modes (broad
background), then the background would become Poissonian. Hence a
$g^{(2)}(0)$-value would range by a factor of 2 (single mode
thermal state vs pure Poisson) for the same $\mu,$ where the
latter is the average number of photons per pulse. And -- sure
enough -- in the presence of technicalities, such as loss, jitter
and uncorrelated noise, exposure time (time window) would matter.

\section{Self-inconsistency of dichotomous quantum observables in the light of random field model of quantum phenomena}

It is obvious that one can ignore our analysis of self-consistency
of quantum phenomena by regarding the problem of the coincidence
clicks as a purely technical problem of the elimination of noise,
i.e., having no fundamental value. This problem can have a
fundamental value only under the assumption that this problem
cannot in principle be solved by improving technologies for
``single photon'' sources and detectors. Such an assumption cannot
be justified within QM. However, there are some logical reasons
supporting this assumption (sections 7, 8). In particular, one can
present some motivations for it by going beyond the quantum
formalism and considering prequantum (classical probabilistic)
models reproducing quantum probabilities. The first reaction to
such a comment would be that, as a consequence of various no-go
theorems, such prequantum models do not exist or if they exist,
they have to be nonlocal as, e.g., Bohmian mechanics. According to
Einstein, we reject such an ambiguous notion as nonlocal realism.
Therefore, we discuss only local prequantum models. Of course,
such models have to be nonrealistic in the Bell's sense, i.e.,
non-objective in our terminology. We remark that Bell's
terminology ``realism'' in connection to the problem of hidden
variables is a bit ambiguous. He definitely discusses realism of
quantum observables expressed in terms of hidden variables.
However, realism can be recovered on the level of hidden
variables, if quantum observables are not expressible in terms of
such additional variables. The notion of (non)objectivity is
related only to quantum observables.

Thus, we want to discuss a non-objective model with hidden
variables. The key point is that such a model will be
self-inconsistent at the level of measurements formulated in the
yes-no logics and the Aristotelian hypothesis of objectivity
constructed on logical atoms. In spite such features as
non-objectivity and self-referentiality which are ``pathological''
in classical world, our model is very natural. In fact, there is
nothing more natural if one wants to arrive to quantum physics by
departing from classical theory. Non-objectivity in the
Aristotelian meaning and self-referentiality on the level of
observations are strange only for classical mechanics of {\it
particles.} And we consider waves, instead of particles. This
approach was originally explored by Schr\"odinger, but later he
gave up. We were able to resolve the problems which pressed
Schr\"odinger to accept the probabilistic interpretation of the
wave function (due to Max Born); in particular, the problem of the
wave modelling of composite systems.

In short, in our model which is known under the name {\it
prequantum classical statistical field theory} (PCSFT)
\cite{19}--\cite{28} quantum systems are symbolic representations
of classical random fields fluctuating at time and space scales
which are essentially finer than quantum labs scales. Such fields
by interacting with detectors of the {\it threshold type} produce
clicks. These clicks are interpreted as quantum events.

In PCSFT, the irradiance of a beam of light is only an indication
of its average state. If we could magnify local states, we should
see a little bit of chaos. At some points, the amplitude of the
waves is well below the average, and at others we get arbitrarily
high spikes. In short, the field is ``clumpy'' at the microscopic
level.

Suppose that we have a point-like detector. When the field crosses
the plane of detection, it might happen that the local amplitude
is close to average or lower. No detection is possible. It can
also happen that we have an amplitude spike followed by several
small crests. Again, the signal does not accumulate above the
threshold and nothing happens. Yet, there is a real probability
that an amplitude spike will continue over several cycles. In this
case, sustained resonance above the threshold will result in a
detection click. Consequently, the pattern of detection is
produced by the low probability of transient ``spikes'' in a
continuous field. It is not true that we have single discrete
entities at the moment and point of detection.

At the level of such events, PCSFT is fundamentally
self-inconsistent. The probability of a coincidence click, i.e.,
matching of two trains of spikes (at the micro-scale) at two
detectors is nonzero, even theoretically. It decreases with
increase of the threshold, but even for very high threshold a
random field can produce matching spikes.

Moreover, one cannot violate Bell's inequality and more generally
to represent quantum compound systems in entangled states by
considering random fields propagating in vacuum (at least in our
model). One has to consider a random {\it background field}  which
is present everywhere (one may call it zero point field or vacuum
fluctuations). This (classical) field contributes into
correlations and, in particular, its presence gives a possibility
to violate Bell's inequality. This field has the random structure
which similar to the one of random fields-signals representing
quantum systems. Hence, a threshold detector ``eats'' energy of
combined spikes, signals combined with the background field.

Non-objectivity of such observables on random fields is a
consequence of self-referentiality, the impossibility in general
to assign say polarization up or down. As a consequence of the
presence of the random background field contributing irreducibly
into threshold detection, the coincidence clicks appear
irrespectively to our manipulations with random field-signals
representing quantum systems.

\section{Conclusion}

Following Bohr, von Weizs\"acker, Brukner, and Zeilinger, we have
analyzed the problem of inconsistency between classical language
description of theoretical quantum phenomena based on
Aristotelian-Russellian logics and experimental structure of these
phenomena which is exhibited first of all in the complementary
structure of quantum experiments (so-called ``wave-particle
duality''). We have presented this problem in very general context
of linguistic solipsism (Wittgenstein, Searle, and Austin) by
emphasizing the role of performative propositions in scientific
theories and, in particular, in QM. Such propositions are in
general self-referent; attempts to use them in combination with
Aristotelian-Russellian logics leads to inconsistency. We have
argued that, nevertheless, it is possible to escape logical
self-inconsistency of quantum performativity by appealing to the
approach based on non-well-founded reality \cite{Schumann3} (as
opposed to the approach based on the objective reality).

In this paper, we pointed out that the problem of
self-inconsistency of QM is even deeper than self-inconsistency
implied by the principle of complementarity. The latter (see
presentation of views of Brukner and Zeilinger in the
Introduction) implies that, for a quantum system $S,$ it is
impossible to assign consistently the truth values to all
propositions about this systems. We found that even the statement
about existence of a quantum system cannot be peacefully embedded
in Aristotelian-Russellian logics. Our argument is based on the
analysis of the experiment on ``photon existence'', measurement of
the coefficient of second order coherence. If positivity of this
coefficient for experiments with the ``single photon state'' is
interpreted as a foundational issue (and not just as a problem of
noise and the state preparation), then the operational definition
of photon as detector's click leads to self-inconsistency of QM,
self-inconsistency of coupling between the notions of QM as a
theoretical formalism and the real experimental situation.

We also stressed the similarity between quantum mechanical and
biological phenomena. Both are characterized by descriptions based
on performative propositions and they are self-inconsistent in the
framework of Aristotelian-Russellian logics. Our discussion on
biological systems is restricted to performativity related to the
principle of complementarity, impossibility to assign consistently
the truth values to all propositions about actions of a biological
system.

An important part of our consideration was about the role of
experimenter's free will in resolving (at least partially)
self-inconsistency of QM; we have pointed to the performative
nature of statements related to ``experimental technicalities''
such as, e.g., the discrimination threshold and time window. We
have also analyzed the possibility to resolve self-inconsistency
of QM by going ``beyond quantum''. So, we have considered a model
of the classical field type reproducing the basic predictions of
QM, the so-called prequantum classical statistical field theory
(PCSFT). By PCSFT objectives, reality can be recovered at the
subquantum level, in spite of non-objectivity of ``reality at the
quantum level.''

\section*{Acknowledgments}

This research is being fulfilled by the support of FP7-ICT-2011-8
and UMO-2012/07/B/HS1/00263 (A. Schumann) and visiting fellowships
(A. Khrennikov) to the Center of Quantum Bio-Informatics (Tokyo
University of Science).  The authors would like to thank K. Svozil, C. Garola,
S. Sozzo, and E. E. Rosinger for critical discussions on logical foundations 
of quantum mechanics and C. Fuchs, M. D' Ariano, C. Brukner and A. Zeilinger for again critical discussions
on the role of information in quantum foundations.

\end{document}